\documentclass[aps,prx,twocolumn,superscriptaddress,showpacs]{revtex4-1}
\usepackage{amsmath,amssymb,graphics,epsfig,epstopdf,color,verbatim,ulem,braket,tabularx}
\usepackage{multirow}
\usepackage[colorlinks,linkcolor=blue,citecolor=blue,urlcolor=blue,bookmarks=false]{hyperref}
\usepackage{listings}
\usepackage{cancel}
\usepackage{mathrsfs}
\usepackage{soul}
\usepackage{color}

\begin{document}

\title{Pairing enhanced by local orbital fluctuations in a model for monolayer FeSe}

\author{Changming Yue}
\email{changming.yue@unifr.ch}
\affiliation{Department of Physics, University of Fribourg, 1700 Fribourg, Switzerland}

\author{Philipp Werner}
\email{philipp.werner@unifr.ch}
\affiliation{Department of Physics, University of Fribourg, 1700 Fribourg, Switzerland}

\begin{abstract}
The pairing mechanism in different classes of correlated materials, including iron based superconductors, is still under debate. For FeSe monolayers, uniform nematic fluctuations have been shown in a lattice Monte Carlo study to play a potentially important role. Here, using dynamical mean field theory calculations for the same model system, we obtain a similar phase diagram and provide an alternative interpretation of the superconductivity in terms of local orbital fluctuations and phase rigidity. Our study clarifies the relation between the superconducting order parameter, superfluid stiffness and orbital fluctuations, and provides a link between the spin/orbital freezing theory of unconventional superconductivity and theoretical works considering the role of nematic fluctuations.
\end{abstract}

\maketitle

\section{Introduction}
Monolayer FeSe grown on SrTiO$_3$  (STO) exhibits superconductivity with a remarkably high superconducting $T_c$ of more than ten times the bulk value \cite{CPL2012_SC_FeSe, NatCom2012_SC_FeSe, He_NatMat_2013_SC_FeSe, Tan_NatMat_2013_SC_FeSe, CPL2014, PRL2016_SC_FeSe, AnuRev2017_FeSe}.
Various theories have been proposed to explain this surprising experimental result,  
as summarized in Ref.~\onlinecite{AnuRev2017_FeSe}. One possibility is a phononic mechanism, involving an
interface-enhancement of the electron-phonon coupling, as suggested in the original paper \cite{CPL2012_SC_FeSe}.
Lee {\it et al.} observed replica bands using angle-resolved photoemission spectroscopy \cite{Lee2014_FeSe_replica}, consistent with a strong coupling between FeSe electrons and STO phonons.
Using Quantum Monte Carlo simulations, Li {\it et al.} \cite{LI_QMC_FeSe} showed that the $T_c$ can be substantially enhanced by introducing an electron-phonon interaction in the model. 

Significant enhancements of $T_c$, relative to bulk FeSe, are however also found in monolayer systems without STO substrate \cite{Shiogai2021, Wen2016, Miyata2015, Tang2016, Lu2015, Sun2015}. This shows that the interface effect is not the only relevant mechanism, and suggests a significant contribution from a purely electronic mechanism. 
Since bulk FeSe shows a nematic transition around 100~K, but no magnetic ordering, an appealing scenario is that the 
$T_c$ in monolayer FeSe is enhanced by a mechanism related to nematic fluctuations. 
In Ref.~\onlinecite{Dumitrescu2016}, Dumitrescu {\it et al.} used lattice Monte Carlo simulations of a two-band model with attractive intra-orbital interactions to reveal a connection between superconductivity and uniform nematic fluctuations, detected through the ${\bf q=0}$ correlation function for the orbital moments. 

The model considered in Ref.~\onlinecite{Dumitrescu2016} has some similarity to multi-orbital Hubbard models with negative Hund coupling $J$ \cite{Koga2015,Steiner2016,Hoshino2017}. The latter have been studied in connection with unconventional superconductivity in the fulleride compounds A$_3$C$_{60}$ \cite{Capone2002,Capone2009,Nomura2015,Hoshino2017,Yue_2021_SC}. There, the pairing can be related to enhanced local orbital fluctuations and an orbital-freezing crossover. As discussed in Ref.~\onlinecite{Steiner2016}, the two-orbital Hubbard model with $J<0$ can be mapped to the model with $J>0$, which connects orbital freezing to spin freezing and hence to the unconventional superconductivity observed in materials ranging from uranium based compounds \cite{Saxena_UGe2_2000,Aoki_URhGe_2001,Huy_UGoGe_2007,Aoki_uranium_2012,Hoshino2015} to cuprates \cite{Werner2016}. With the aim of a unified description of unconventional superconductivity in mind, it is thus interesting to look at the previously studied model for FeSe monolayers from an orbital-freezing perspective.

Here, we solve the model of Refs.~\cite{Dumitrescu2016,Yamase2013} using single-site dynamical mean field theory (DMFT) \cite{Georges1996} and show that this approximation essentially reproduces the phase diagram established by lattice Monte Carlo simulations in Ref.~\onlinecite{Dumitrescu2016}. Instead of ${\bf q=0}$ fluctuations, we focus on {\it local} orbital fluctuations and ask to what extent these fluctuations contribute to the pairing. We will show that in the regime of weak-to-moderate bare couplings, the interactions induced by local orbital fluctuations play the dominant role in the pairing (as in the case of fullerides), while in the doped Mott regime, the bare attraction becomes more relevant. 
We will comment on the realistic range for the bare interaction, which is below the critical value for a paired Mott state. 

The paper is organized as follows. In Sec.~\ref{sec:model} we introduce the effective two-band Hubbard model for monolayer FeSe, and the DMFT method used to solve it. In Sec.~\ref{sec:results}, we show the DMFT phase diagram and connect the superconducting order parameter to the effective attractive interaction, orbital fluctuations and the superfluid stiffness. Section~\ref{sec:conclusions} contains a summary and conclusions.       

\begin{figure}[h!]
\includegraphics[clip,width=3.4in,angle=0]{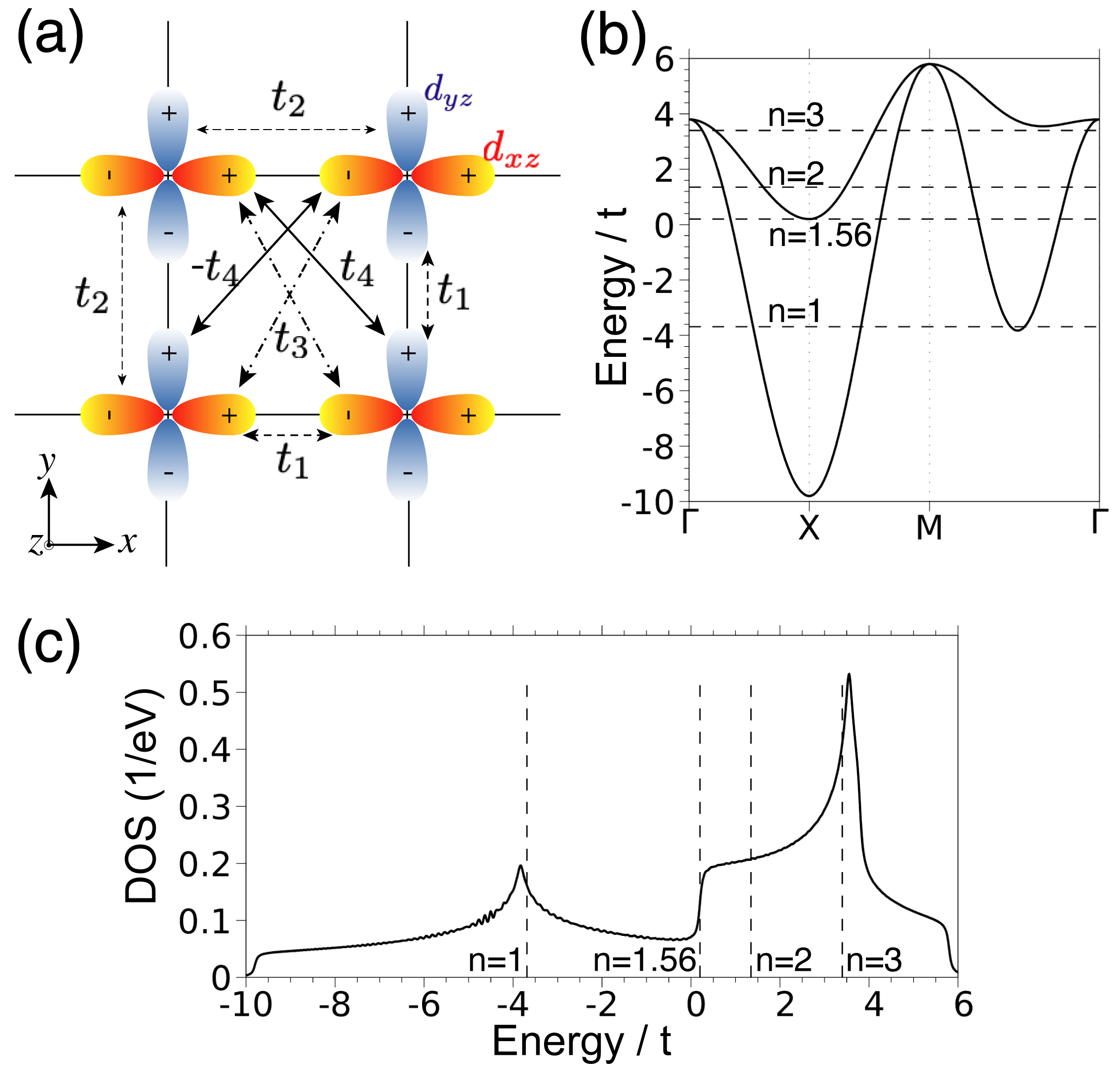}
\caption{
(a) The two-band tight-binding model (adapted from Ref.~\onlinecite{Dumitrescu2016}) on the square lattice, with orbitals $d_{xz}$ and $d_{yz}$ schematically shown in red and blue.  The hoppings are $t_1=-1.0t$, $t_2=1.5t$, $t_3=-1.2t$, $t_4=-0.95t$.  (b) The band structure along the indicated path in the Brillouin zone. (c) The density of states (DOS).  The black dashed lines in (b) and (c) mark the chemical potentials corresponding to $n=1$, $n=1.56$, $n=2$ (half-filling) and $n=3$.}
\label{fig:fig1_TB}
\end{figure}

\section{Model and Method}
\label{sec:model}

For the modeling of monolayer FeSe, we follow Refs.~\onlinecite{Dumitrescu2016,Yamase2013} 
and consider a two-band Hubbard model on the square lattice, 
\begin{align}
H=&-\sum_{\mathbf{i,j}, a, b, \sigma}\left(t_{i j}^{ab} c_{i a \sigma}^{\dagger} c_{\mathbf{j} b \sigma}+\text { H.c.}\right)-\mu \sum_{\mathbf{i}, a} n_{\mathbf{i}, a} \nonumber\\ 
&-\frac{g}{2} \sum_{\mathbf{i}}\left(n_{\mathbf{i}, x z}-n_{\mathbf{i}, y z}\right)^{2}, 
\label{eq:model}
\end{align}
with orbitals of $d_{xz}$, $d_{yz}$ character, as illustrated in Fig.~\ref{fig:fig1_TB}(a). 
Here, $\mathbf{i,j}$ label the lattice sites, $a,b=d_{xz}, d_{yz}$ the orbitals, and $\sigma=\uparrow,\downarrow$ the spin, respectively. 
The first term in Eq.~(\ref{eq:model}) is the non-interacting tight-binding model $H_0$, for which the nonzero hopping terms $t_{i j}^{a b}$ are 
shown by the arrows in Fig.~\ref{fig:fig1_TB}(a). The second term in Eq.~(\ref{eq:model}), with the number operator defined as 
$n_{ia}=n_{ia\uparrow}+n_{ia\downarrow}$, allows to adjust the filling by varying the chemical potential $\mu$.  For $g>0$, the last term penalizes an equal occupation of the two orbitals on a given site and favors the formation of an orbital moment. Such an interaction term has been argued in Ref.~\onlinecite{Dumitrescu2016} to originate from Fe-ion oscillations and electron-phonon coupling \cite{Onari_PRL_2015}, although it should be noted that the work of Kontani and Onari did not consider a regime of bare attractive interactions. We use here the (oversimplified) model of Ref.~\onlinecite{Dumitrescu2016} because our goal is to connect the discussion on nematicity-induced pairing to that on spin/orbital freezing \cite{Hoshino2015,Steiner2016,Hoshino2017}. 

In momentum (${\bf k}$) space and in the Nambu-formalism, $H_0$ can be expressed as 
\begin{equation}
H_{0}=\sum_{\boldsymbol{k}}\left[\begin{array}{cc}
\Psi_{\boldsymbol{k},\uparrow}^{\dagger} & \Psi_{-\boldsymbol{k},\downarrow}\end{array}\right]\left[\begin{array}{cc}
H_{0}({\bf k}) & 0\\
0 & H_{0}(-{\bf k})^{T}
\end{array}\right]\left[\begin{array}{c}
\Psi_{\boldsymbol{k},\uparrow}\\
\Psi_{-\boldsymbol{k},\downarrow}^{\dagger}
\end{array}\right]
\end{equation}
where the Nambu spinors are
$[\begin{array}{cc}
\Psi_{\boldsymbol{k},\uparrow}^{\dagger} & \Psi_{-\boldsymbol{k},\downarrow}\end{array}]=[\begin{array}{cccc}
c_{{\bf k},1\uparrow}^{\dagger} & c_{{\bf k},2\uparrow}^{\dagger} & c_{-{\bf k},1\downarrow} & c_{-{\bf k}2\downarrow}\end{array}]$.
Here, $H_{0}({\bf k})=\left[\begin{array}{cc}
\epsilon_{{\bf k}}^{11} & \epsilon_{{\bf k}}^{12}\\
\epsilon_{{\bf k}}^{12} & \epsilon_{{\bf k}}^{22}
\end{array}\right]$
is a $2\times 2$ matrix with the elements
$\epsilon_{\bf k}^{11}=-2t_{1}\cos k_{x}-2t_{2}\cos k_{y}-4t_{3}\cos k_{x}\cos k_{y}$,
$\epsilon_{\bf k}^{22}=-2t_{2}\cos k_{x}-2t_{1}\cos k_{y}-4t_{3}\cos k_{x}\cos k_{y}$, 
and $\epsilon_{\bf k}^{12}=-4t_{4}\sin k_{x}\sin k_{y}$. For the hopping amplitudes, we use $t_1=-1.0t$, $t_2=1.5t$, $t_3=-1.2t$, $t_4=-0.95t$,
which are expressed in units of $t = 100$ meV \cite{Hung2012,Raghu2008,Yao2009FeSemodel}.
Since $\epsilon_{\bf k}^{\alpha\beta}$ is even in ${\bf k}$ 
we have 
$H_{0}(-{\bf k})^{T}=H_{0}({\bf k})$.
The band structure and density of states (DOS) of $H_{0}({\bf k})$ are shown in Fig.~\ref{fig:fig1_TB}(b) and Fig.~\ref{fig:fig1_TB}(c), respectively. Clearly, there is no particle-hole symmetry in the tight-binding model. 
The chemical potentials associated with filling $n=1$, $2$, and $3$ are indicated in the band structure and in the DOS. In addition, we highlight the filling $n=1.56$ corresponding to the lower edge of the upper band, since the jump in the DOS at this value leaves clear traces in the results presented in Sec.~\ref{sec:results}. Because of the broad band with weak van Hove singularity near $n=1$, and the more narrow band with prominent van Hove singularity near $n=3$, we expect stronger correlation effects on the electron doped side than on the hole doped side of the half-filled ($n=2$) system. 

The interaction term $H_{\mathrm{int}}$ (third term in Eq.~(\ref{eq:model})) can be decomposed using $g\equiv -U>0$ into a chemical potential shift and intra/inter-orbital density-density interaction terms,
\begin{align}
H_{\mathrm{int}}= \frac{U}{2}\sum_{i\alpha\sigma} n_{i\alpha\sigma}+U\sum_{i\alpha}n_{i\alpha\uparrow}n_{i\alpha\downarrow}-U\sum_{i\sigma\sigma^{\prime}}n_{i1\sigma}n_{i2\sigma^{\prime}}.
\label{eq:Hint}
\end{align}
While the bare interaction parameters estimated for $d$-electron models of iron pnictides are repulsive \cite{Anisimov2009,Miyake2010}, it has been argued in Ref.~\onlinecite{Onari_PRL_2015} that a moderate local electron-phonon coupling substantially screens these interactions and results in a situation where orbital fluctuations, rather than spin fluctuations, play a dominant role. Within our phenomenological description, $U<0$ allows to mimic this situation, but one should keep in mind that large attractive on-site interactions $U$ are unrealistic.   
With $U<0$, $H_{\mathrm{int}}$ favors intra-orbital spin-singlet pairing \cite{Koga2015} and model \eqref{eq:model} becomes similar to a two-orbital Hubbard model with negative Hund coupling $J$. The difference is that the inter-orbital same-spin and opposite-spin interactions are equal, which is not the case in the usual Kanamori model with Hund coupling, but at the qualitative level, we can expect similar low-energy physics. 

We solve the correlated lattice system within the framework of DMFT \cite{Georges1996}, where the lattice model is mapped onto a self-consistently determined quantum impurity model. 
This two-orbital impurity model is solved using the hybridization-expansion continuous-time quantum Monte-Carlo (CT-HYB) algorithm \cite{Werner2006,Werner2006b,Gull2011}. 
The hybridization function is diagonal in orbital space because $\epsilon_{\bf k}^{12}$ satisfies $\epsilon_{(k_{x},k_{y})}^{12}=-\epsilon_{(-k_{x},k_{y})}^{12}=-\epsilon_{(k_{x},-k_{y})}^{12}$,
which leads to an orbital-diagonal local Green's function. 
We use here a Nambu implementation of the DMFT loop, as described in Refs.~\onlinecite{Georges1996,Koga2015}, in order to treat the superconducting phase.
To reduce the noise in the impurity self-energy, we employ (symmetric) improved estimators \cite{Hafermann2012,Kaufmann2019}.
To map out the phase diagram, we allow for orbital and sublattice symmetry breaking (ferro- and antiferro-orbital order, as well as charge order), but in the study of the superconducting state we will suppress these orders. 

The results are shown for temperature $T=t/8$, which corresponds to 12.5 meV or 145 K (same as in Ref.~\onlinecite{Dumitrescu2016}), unless otherwise noted, and we use $t = 100$ meV as the unit of energy. 

\begin{figure}[t]
\includegraphics[clip,width=3.4in,angle=0]{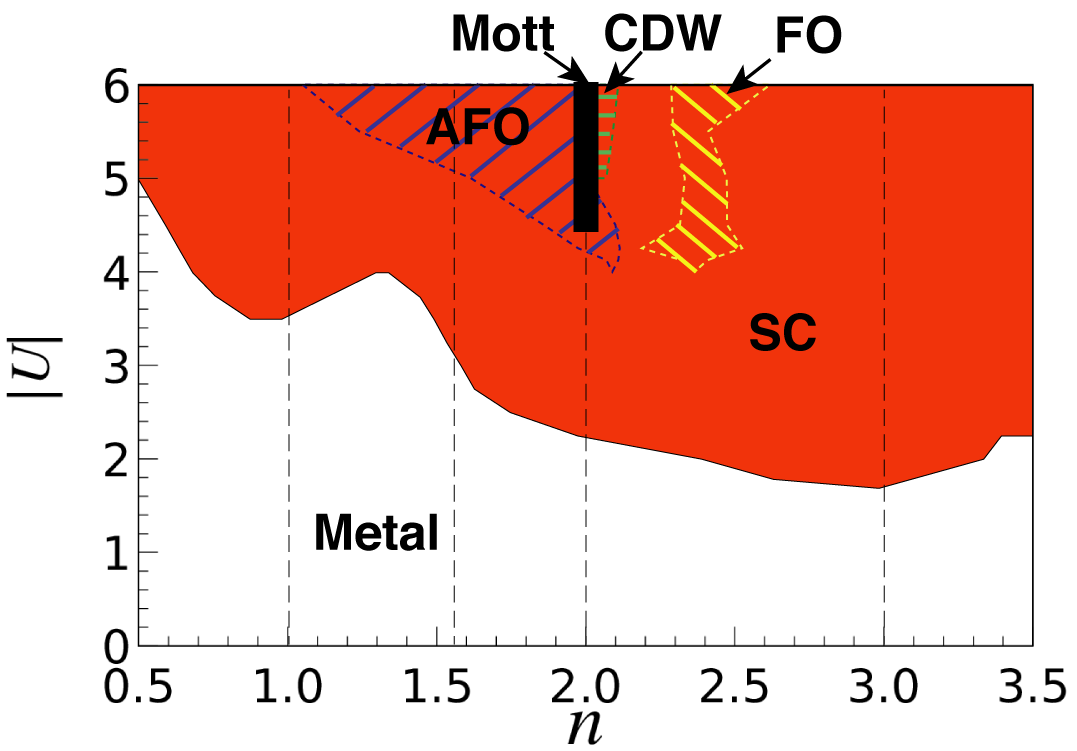}
\caption{
DMFT phase diagram of model (\ref{eq:model}) in the space of interaction $|U|$ $(=g)$ and total filling $n$, at temperature $T=1/8$. 
The white region indicates the normal metallic phase, and the red region the superconducting (SC) phase with order parameter $\Delta \ge 0.01$. 
The orbital and/or sublattice symmetry breaking phases AFO, CDW, and FO are sketched by the blue, green and yellow hashed regions, respectively.
The thin dashed lines mark the same fillings as in Fig.~\ref{fig:fig1_TB}.
}
\label{fig:phase_diagram}
\end{figure}

\begin{figure}[h!]
\includegraphics[clip,width=3.2in,angle=0]{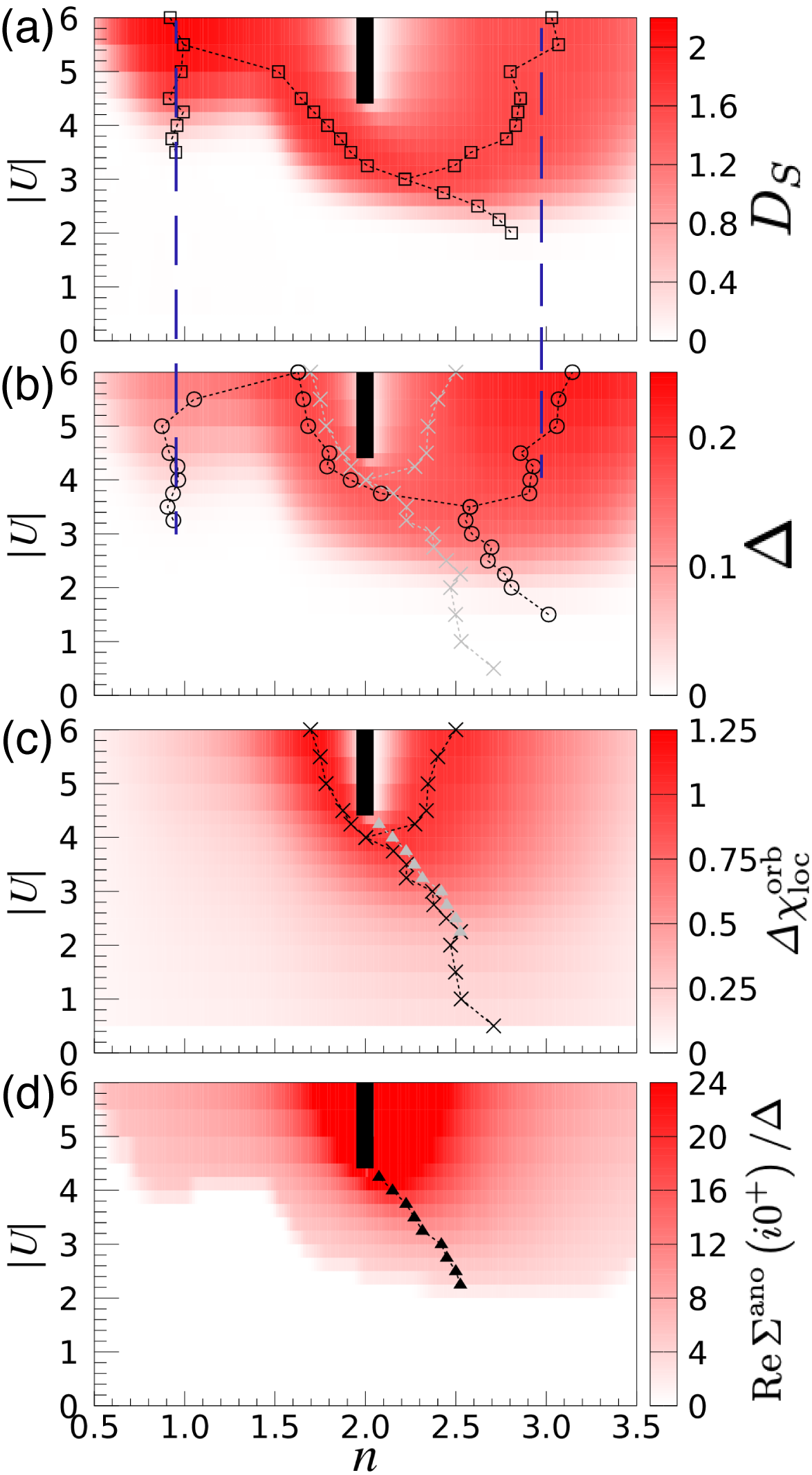}
\caption{
Filling $n$ and interaction $|U|$ dependence of (a) the superfluid stiffness $D_S$, 
(b) the SC order parameter $\Delta$, (c) the local orbital fluctuations ${\it \Delta} \chi_{\mathrm{loc}}^{\mathrm{orb}}$
and (d) the effective interaction $\mathrm{Re}\Sigma^{\mathrm{ano}}(i0^+)/\Delta$ (white-red color map) for orbital symmetric phases at $T=1/8$.
The thick black bars in all panels indicates the paired Mott insulating phase. 
The black squares in (a), circles in (b), crosses in (c) and triangles in (d) mark the corresponding peak positions at a fixed 
interaction $|U|$ along the axis of filling $n$.
For the sake of easier comparison, we also indicate the peak positions from panel (d) in panel (c), and similarly those from panel (c) in panel (b).
The blue dashed lines link the peak positions in $\Delta$ and $D_S$. 
Due to a large error bar in determining the SC phase near $n=2$, we truncate the data in (d) with a cutoff $\Delta>0.03$. 
The stiffness is $D_S=D_S^{xx}$ with units $e^2/\hbar^2$.
}
\label{fig:fig3_colormap}
\end{figure}

\begin{figure}[h!]
\includegraphics[clip,width=3.4in,angle=0]{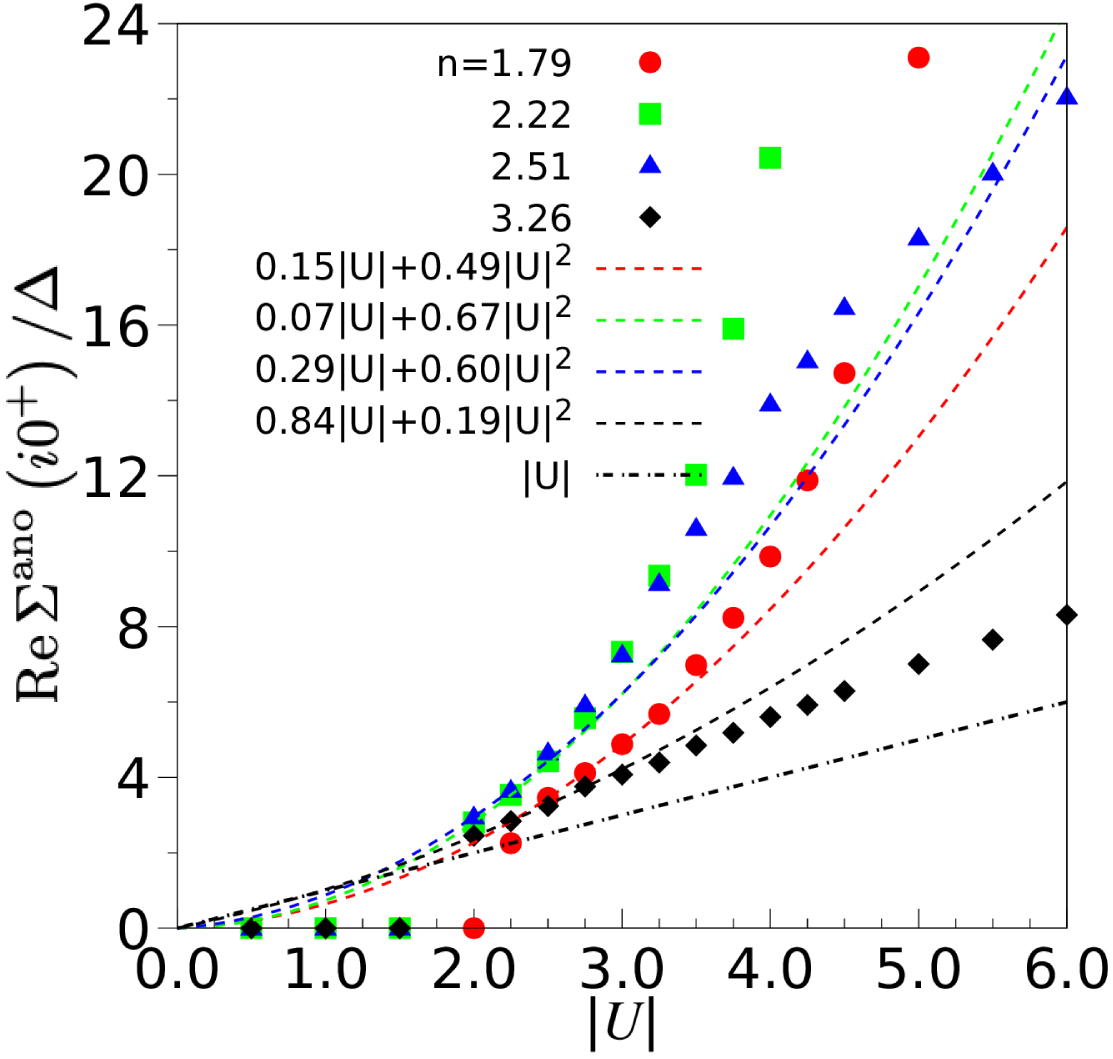}
\caption{
Effective interaction $|U_{\mathrm{eff}}^\text{DMFT}|=|\mathrm{Re}\Sigma^{\mathrm{ano}}(i0^+)/\Delta|$ as a function of bare interaction $|U|$ (points) and parabolic fits $a|U| + b |U|^2$ to the small-$|U|$ data (dashed lines) for indicated fillings.  The black dot-dashed line plots $|U|$.}
\label{fig:fig6_Ueff}
\end{figure}

\section{Results}
\label{sec:results}

\subsection{Phase diagram and orbital fluctuations}

The main results of our study are summarized in Figs.~\ref{fig:phase_diagram} and \ref{fig:fig3_colormap}. The phase diagram with superconducting (SC), antiferro-orbital order (AFO), ferro-orbital order (FO) and charge density wave (CDW) phases is shown in Fig.~\ref{fig:phase_diagram}. Here, the thick black line indicates the (paired) Mott phase at $n=2$ in the system with suppressed electronic orders and attractive interaction $U\lesssim -4.4=U^\text{Mott}_c$. The appearance of AFO order near half-filling and FO order in the doped system can be understood by looking at the generic DMFT phase diagram of the two-orbital Hubbard model with $J>0$ in Ref.~\onlinecite{Hoshino2016} and considering the fact that switching $J>0\rightarrow J<0$ maps ferromagnetism to FO and anti-ferromagnetism to AFO order (as well as spin-triplet SC to spin-singlet SC) \cite{Steiner2016}, and that our system is qualitatively similar to the $J<0$ case. Because ferromagnetism (and hence FO order) appears only at strong coupling \cite{Hoshino2016}, we detect FO only on the electron-doped side. The appearance of a CDW in the half-filled Mott system is similar to what one finds in the attractive single-band Hubbard model, where SC and CDW coexist at half-filling \cite{Scalettar1989}. The strong asymmetry of the phase diagram with respect to $n=2$ appears because of the strongly asymmetric DOS. 

The superconducting order parameter $\Delta=\langle c_{1\sigma\uparrow} c_{1\sigma\downarrow}\rangle=\langle c_{2\sigma\uparrow} c_{2\sigma\downarrow}\rangle$ at $T=1/8$ in states with suppressed sublattice and orbital symmetry breaking is plotted in panel (b) of Fig.~\ref{fig:fig3_colormap}. These results demonstrate the much stronger pairing near $n\approx 3$, compared to $n\approx 1$, and a substantial decrease in the order parameter below the step in the DOS ($n\lesssim 1.56$), as one may expect based on the different correlation strengths in the respective filling regimes. Our results are similar to the lattice Monte Carlo results reported in Ref.~\onlinecite{Dumitrescu2016} as far as the stability regions of the SC and AFO phases are concerned, and also with regard to the filling dependence of the order parameter.  What is different is that the lattice simulations have not detected any FO and CDW instabilities. Here, we have to note that lattice simulations on relatively small lattices cannot easily distinguish short-range correlations from long-range oder, while DMFT treats these orders at the mean-field level and has a tendency to overestimate their stability region. In the following, we will suppress AFO, FO and CDW order to investigate the properties of the SC state and connect the latter to orbital fluctuations. 

First, it should be noted that the appearance of local singlet pairing in a model with a bare on-site attractive interaction is of course expected. However, as noted in  Ref.~\onlinecite{Dumitrescu2016}, the SC phase in a mean-field treatment of the model appears at rather large interaction, $|U|>6$ at $T=1/8$, so that the superconducting states revealed in Fig.~\ref{fig:phase_diagram} and Fig.~\ref{fig:fig3_colormap}(b) must be stabilized, or at least enhanced, by an additional source of attractive interactions. Here, we focus on the role of {\it local} orbital fluctuations. 

Following Refs.~\onlinecite{Suga2013,Hoshino2015} we can, based on a weak-coupling picture, derive an effective interaction $U_\text{eff}$ which takes into account the leading correction from bubble diagrams. In Appendix~\ref{App_U} we show that for the current model, with $g=-U$, one finds $|{U}_\text{eff}^\text{bubble}|=|U|+2|U|^{2}\chi_{\mathrm{loc}}^\text{orb}(\omega=0)$, where $\chi_{\mathrm{loc}}^\text{orb}(\omega=0)$ is the static value of the Fourier transform of the
orbital-orbital correlation function 
$\chi_{\mathrm{loc}}^{\mathrm{orb}}(\tau)=\tfrac 14 \langle (n_1(\tau)-n_2(\tau))(n_1(0)-n_2(0))\rangle$.
In the strongly correlated regime, the orbital moment can freeze \cite{Steiner2016} and it is more natural to replace $\chi_{\mathrm{loc}}^\text{orb}(\omega=0)$ by the ``fluctuating contribution" ${\it \Delta}\chi_{\mathrm{loc}}^{\mathrm{orb}}=\int_0^\beta \chi_{\mathrm{loc}}^{\mathrm{orb}}(\tau) \mathrm{d}\tau-\beta \chi_{\mathrm{loc}}^{\mathrm{orb}}(\beta/2)$. Based on these arguments, we expect that the attractive interaction is enhanced as 
\begin{equation}
|{U}_\text{eff}^\text{bubble}|=|U|+2|U|^{2}{\it \Delta}\chi_{\mathrm{loc}}^{\mathrm{orb}}\label{Ueff} ,
\end{equation}
with a correction term that is proportional to the square of the bare $|U|$, at least in the weak-coupling regime. 

As discussed in the fulleride context in Ref.~\onlinecite{Yue_2021_SC}, the effective attractive interaction in the SC state can be measured by computing the ratio between the real part of the anomalous self-energy  in the static limit $\mathrm{Re}\Sigma^{\mathrm{ano}}(i0^+)$, and the order parameter $\Delta$. This provides a way of testing the qualitative prediction in Eq.~\eqref{Ueff}. Figure~\ref{fig:fig6_Ueff} plots $\mathrm{Re}\Sigma^{\mathrm{ano}}(i0^+)/\Delta\equiv U_\text{eff}^\text{DMFT}$ as a function of $|U|$ for different fillings, together with a fit to a linear plus quadratic function. We see that $|U_\text{eff}^\text{DMFT}|\ge |U|$ and that at least in the small-$|U|$ regime, where the simple bubble-estimate (\ref{Ueff}) is meaningful, the enhancement of the attractive interaction is approximately quadratic \footnote{Note that the prefactor of the linear term is $<1$, as it was the case in Ref.~\onlinecite{Yue_2021_SC}. This means that the bubble calculation works at a qualitative level, for properly renormalized interaction parameters.}. This provides direct evidence for an enhancement of the pairing interaction, and hence SC, by local orbital fluctuations. 

To further investigate the link between $\mathrm{Re}\Sigma^{\mathrm{ano}}(i0^+)/\Delta$ and ${\it \Delta} \chi_{\mathrm{loc}}^{\mathrm{orb}}$, we show these quantities as intensity plots in panels (c) and (d) of Fig.~\ref{fig:fig3_colormap}. We furthermore show by the dashed black line with crosses in (c) the peak values of ${\it \Delta} \chi_\text{loc}^\text{orb}$ and by the black dashed line with triangles in (d) the location of the maxima in $\mathrm{Re}\Sigma^{\mathrm{ano}}(i0^+)/\Delta$ for $|U| \lesssim U_c^\text{Mott}$. We also reproduce the maxima from panel (d) by the gray line with triangles in panel (c). One finds that for interactions smaller than the  $U_c^\text{Mott}$ of the $n=2$ paired Mott insulating state, there is an almost perfect match between the maxima in the local orbital fluctuations and the maxima in the effective attractive interaction, which further supports the picture of pairing induced by local orbital fluctuations. In this regime, the situation is hence very similar to the fulleride systems discussed in Refs.~\onlinecite{Hoshino2017,Yue_2021_SC}, even though in the present case we have an attractive bare interaction, while the bare interactions are repulsive (but $J<0$, similar to here) in the fulleride case. 

\begin{figure}[h!]
\includegraphics[clip,width=3.2in,angle=0]{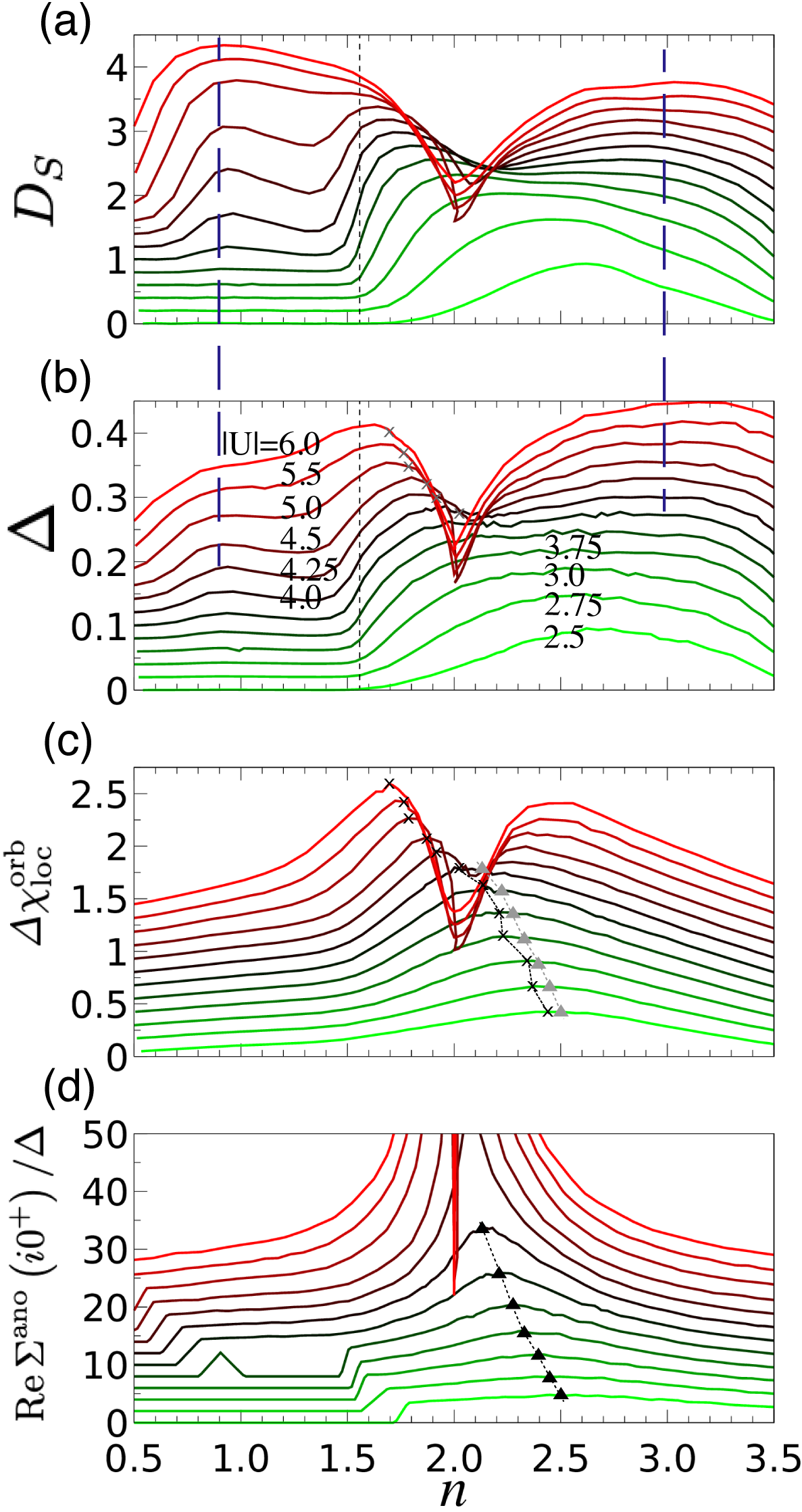}
\caption{
Filling ($n$) dependence of (a) the superfluid stiffness $D_S$, 
(b) the SC order parameter $\Delta$, (c) the local orbital fluctuations ${\it \Delta} \chi_{\mathrm{loc}}^{\mathrm{orb}}$
and (d) the effective interaction $\mathrm{Re}\Sigma^{\mathrm{ano}}(i0^+)/\Delta$ at a series of $|U|$ values as indicated in panel (b).
The black crosses in (c) and triangles in (d) mark the corresponding peak positions.
For a better comparison between the peak positions, we also indicate the peak positions from panel (d) in panel (c) by the gray triangles. Similarly, we reproduce the maxima of (c) by the gray crosses in (b) on the hole-doped side. Due to large error bars in determining the SC phase near $n=2$, we truncate the data in panel (d) with a cut-off $\Delta>0.03$. 
The curves in panel (a) [(b,c,d)] are shifted by multiples of 0.2 [$(0.02,0.125,2)$] along the vertical axis for a better presentation.
The thin dashed lines in panels (a,b) mark the filling $n=1.56$.
}
\label{fig:fig3_linecuts_new}
\end{figure}

To make the connection between the enhanced local orbital fluctuations and the effective attractive interaction even more clear, we plot in panels (c) and (d) of Fig.~\ref{fig:fig3_linecuts_new} several cuts at fixed $|U|$ values. Again, the positions of the maxima in ${\it \Delta}\chi_\text{loc}^\text{orb}$ are indicated by crosses, and the maxima in the measured effective attractive interaction by triangles, and we reproduce the maxima in (d) by the gray triangles in panel (c). 

As discussed in several previous works \cite{Toschi2005,Simard2019,Yue_2021_SC}, to understand how the order parameter depends on the filling or (effective) interaction, one also needs to consider the superfluid stiffness $D_S$, which we can compute from the Nambu Green's functions as explained in Appendix~\ref{App_stiffness}. $D_S$ is plotted in panel (a) of Figs.~\ref{fig:fig3_colormap} and \ref{fig:fig3_linecuts_new}. The stiffness gets smaller for more strongly correlated systems, as one can see in Fig.~\ref{fig:fig3_linecuts_new} from the correlation between the peak in $\mathrm{Re}\Sigma^{\mathrm{ano}}(i0^+)/\Delta$ and the dip in $D_S$ near $n\approx 2$ and $|U|\gtrsim |U_c^\text{Mott}|$, or by noticing the larger value of $D_S$ on the hole doped side, compared to the electron doped side for large $|U|$. While the situation for  $|U| \lesssim |U_c^\text{Mott}|$ is complicated, and the maxima in $\Delta$ seem to correlate both with the maxima in ${\it \Delta} \chi_{\mathrm{loc}}^{\mathrm{orb}}$ and those in $D_S$, for larger $|U|$ and on the electron-doped side, we clearly find that the maximum in $\Delta$ appears in the filling region ($n\approx 3$) where the stiffness is maximal. The same is true for the strongly hole-doped system near $n\approx 1$. 
At these doping levels, orbital freezing is no longer effective and hence there is no longer a match between the maxima in the local orbital fluctuations and the maxima in $\Delta$. This indicates that in the large-$|U|$ and large-doping regime, the pairing gets dominated by the bare attractive interaction, rather than by the fluctuation-induced retarded effective attraction. 
On the other hand, in the weakly doped large-$U$ regime, where the orbital-freezing crossover takes place, there is still a good correlation between the maxima in the orbital fluctuations (see crosses in Fig.~\ref{fig:fig3_linecuts_new}(b)) and the maxima in $\Delta$, and similarly, we may interpret the fast rise of $\Delta$ on the electron-doped side as an effect of orbital-fluctuation-enhanced pairing. In the latter regime, we also note the apparent connection between the maxima in $\Delta\chi^\text{orb}_\text{loc}$ and the FO instability (compare Figs.~\ref{fig:phase_diagram} and  Fig.~\ref{fig:fig3_colormap}(c)). The rapid decrease in $D_S$ and $\Delta$ with hole doping around $n=1.56$ (marked by the dashed line in Fig.~\ref{fig:fig3_linecuts_new}(a,b)) is related to the jump in the DOS at the lower edge of the upper band.  

\subsection{Analysis of the spectral functions}

We next investigate the real-frequency spectra of the (orbital- and spin-symmetric) single-particle normal
Green's functions $G_{\alpha\sigma}(\tau)=-\langle T_\tau c_{\alpha\sigma}(\tau)c^\dagger_{\alpha\sigma}(0)\rangle$ and the anomalous Green's functions $F_\alpha(\tau)=-\langle T_\tau c_{\alpha\uparrow}(\tau)c_{\alpha\downarrow}(0)\rangle$, and compare them to the spectral function of the orbital correlation function $\chi_\text{loc}^\text{orb}(\tau)$. While the normal spectral function $A^\text{nor}(\omega)=-\frac{1}{\pi}\text{Im}G(\omega)$ is positive, the anomalous one $A^\text{ano}(\omega)=-\frac{1}{\pi}\text{Im}F(\omega)$ may have negative spectral weight. For the calculation of the latter, we employ the maximum
entropy analytic continuation \cite{JARRELL1996133} of auxiliary Green's
functions (MaxEntAux) \cite{PRB_MaxEntAux} with positive-definite spectral weight. The idea is to introduce the
operators $\hat{a}_{\alpha}=\frac{1}{\sqrt{2}}[c_{\alpha\uparrow}+c_{\alpha\downarrow}^{\dagger}]$
and $\hat{b}_{\alpha}=\frac{1}{\sqrt{2}}[c_{\alpha\uparrow}-c_{\alpha\downarrow}^{\dagger}]$, as well as the two auxiliary Green's functions

\begin{align}
G_{\alpha}^{a,\mathrm{aux}}(\tau) & \equiv-\langle\mathcal{T}_{\tau}\hat{a}_{\alpha}(\tau)\hat{a}_{\alpha}^{\dagger}(0)\rangle \nonumber \\
 & =\frac{1}{2}\left[G_{\alpha\uparrow}(\tau)-G_{\alpha\downarrow}(-\tau)+2F_{\alpha}(\tau)\right],
\end{align}
and 
\begin{align}
G_{\alpha}^{b,\mathrm{aux}}(\tau) & \equiv-\langle\mathcal{T}_{\tau}\hat{b}_{\alpha}(\tau)\hat{b}_{\alpha}^{\dagger}(0)\rangle \nonumber \\
 & =\frac{1}{2}\left[G_{\alpha\uparrow}(\tau)-G_{\alpha\downarrow}(-\tau)-2F_{\alpha}(\tau)\right].
\end{align}
From the corresponding spectra, the spectral function of the anomalous Green's function can be extracted
as 
\begin{align}
A_{\alpha}^{\mathrm{ano}}(\omega) &=-\frac{1}{\pi}\mathrm{Im}F_{\alpha}(\omega) 
=\frac{1}{2}\left[A_{\alpha}^{a,\mathrm{aux}}(\omega)-A_{\alpha}^{b,\mathrm{aux}}(\omega)\right].
\end{align}

\begin{figure}[t]
\includegraphics[clip,width=0.42\paperwidth,angle=0]{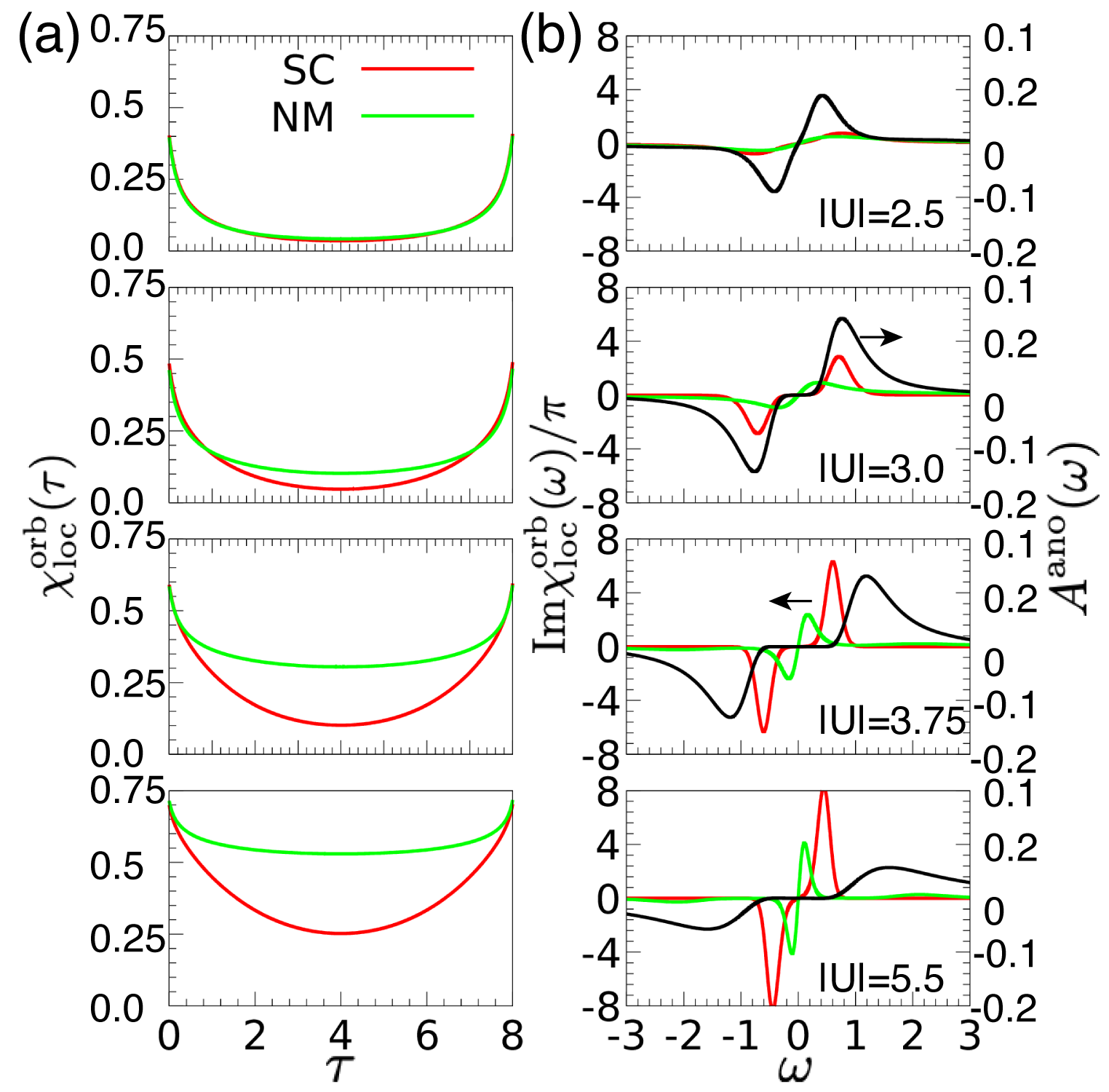}
\caption{
(a) Orbital correlation function in the normal (green) and superconducting (red) state, plotted on the imaginary-time axis.
(b) The spectra of the local orbital correlation function (red, green) and the anomalous Green's function (black).
The red (green) curves are for the SC (normal) phase. 
The filling is $\approx 2.33$ for all subpanels and the interaction strengths are indicated in panel (b).
}
\label{fig:fig5_ChiwFw_Chitau}
\end{figure}

Figure~\ref{fig:fig5_ChiwFw_Chitau} shows $A^\text{ano}(\omega)$ for $n=2.3$ and indicated values of $|U|$ by the black lines in panel (b). We see that with increasing $|U|$ the peak in the spectrum shifts to higher energies and broadens. An interesting question concerns the relation of this peak to the characteristic energy of the orbital fluctuations. In the case of A$_3$C$_{60}$, we showed that (i) the bosonic fluctuations are enhanced in the SC phase, compared to the normal phase, and (ii) on the strong coupling side of the $T_c$ dome (``orbital-frozen" regime) the energies of the peaks in $A^\text{ano}$ and $\text{Im}\chi^\text{orb}_\text{loc}/\pi$ approximately match, because the transition into the SC state melts the orbital freezing and lifts the energy scale of the orbital fluctuations up to that of the pairing fluctuations.  

In panel (a) of Fig.~\ref{fig:fig5_ChiwFw_Chitau} we plot the orbital correlation functions in the normal metal state (green) and in the SC state (red), while the corresponding lines in panel (b) show the bosonic spectral functions $\text{Im}\chi^\text{orb}_\text{loc}(\omega)/\pi$. At the qualitative level, we find the same effect as previously discussed for the repulsively interacting fulleride model, namely that  the orbital freezing, which manifests itself at large $|U|$ by the slow decay of the orbital correlation function, partially melts in the superconducting state, which results in an enhancement of the peak in the spectral function and a shift of the peak to higher energy. In the weak-coupling regime, the bosonic energy scale is higher than the fermionic one, while for strong couplings, in the normal phase, it is lower, again in qualitative agreement with the results of Ref.~\onlinecite{Yue_2021_SC}. However, there is no lock-in between the bosonic and fermionic energy scales in the large-$|U|$ superconducting state, even though the former is clearly increased compared to the normal phase. The missing lock-in phenomenon in the strongly correlated electron-doped compound is another indication that the pairing in this regime occurs not only because of  fluctuation-mediated retarded interactions, but to a significant extent because of the attractive bare interaction. This distinguishes model \eqref{eq:model} from the fulleride systems with purely repulsive bare interactions.  

\begin{figure}[t]
\includegraphics[clip,width=3.4in,angle=0]{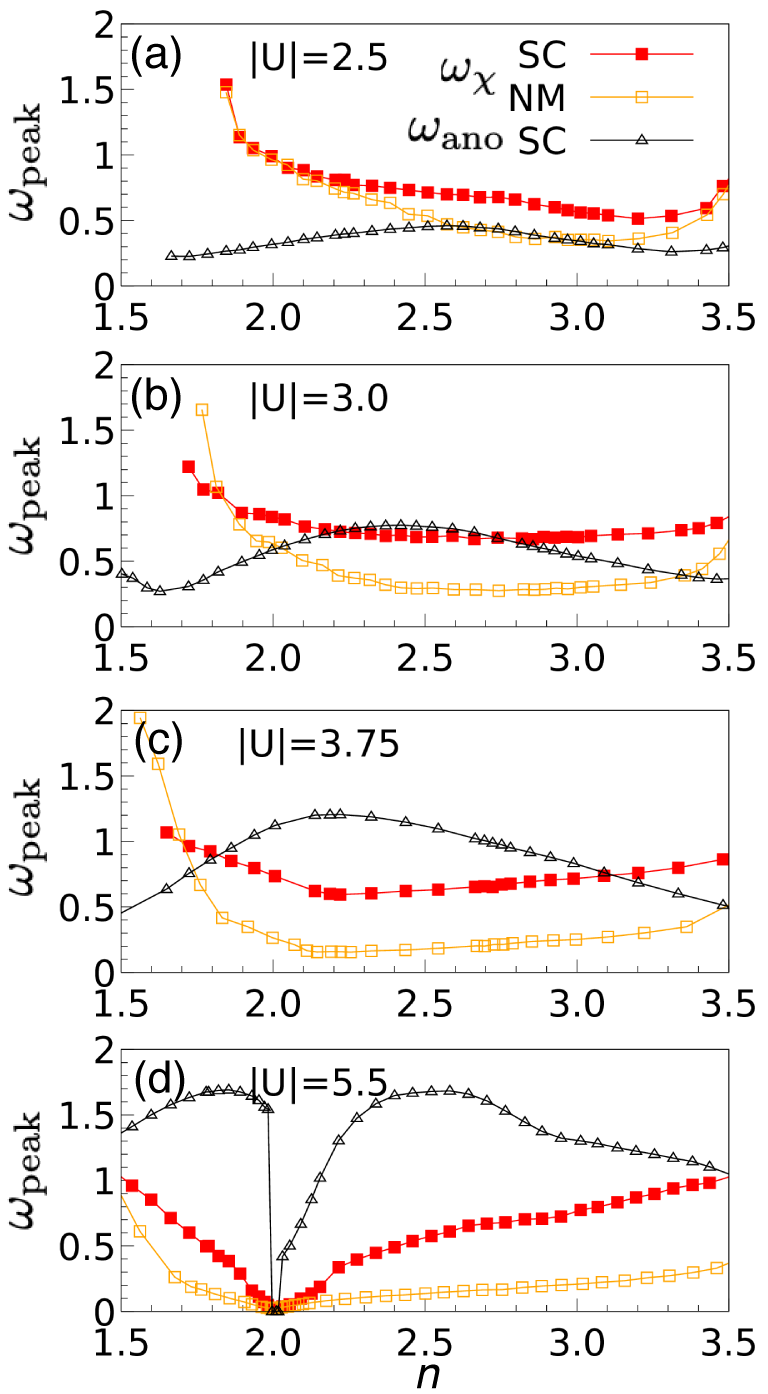}
\caption{
Energy of the peak in $A^{\mathrm{ano}}(\omega)$ (empty triangles) and of the peak in
$\mathrm{Im} \chi_{\mathrm{orb}}^{\mathrm{loc}}(\omega)$ for the SC phase (red solid squares)
and the normal metal (NM) phase (orange empty squares), at (a) $|U|$=2.5, (b) $|U|$=3.0, (c) $|U|$=3.75, and (d) $|U|$=5.5. 
}
\label{fig:fig4_peakpos}
\end{figure}

 A systematic analysis of the energies of the main peaks in $A^\text{nor}$, $A^\text{ano}$ and $\text{Im}\chi_\text{loc}^\text{orb}/\pi$ as a function of filling and bare $|U|$ yields the curves shown in Fig.~\ref{fig:fig4_peakpos}. These results confirm the general trend of an increasing (decreasing) characteristic energy in $A^\text{ano}$ ($\text{Im}\chi_\text{loc}^\text{orb}/\pi$) with increasing $|U|$, the significant increase in the bosonic energy when switching from the normal to the superconducting phase, especially for larger $|U|$, as well as the absence of a lock-in between the peaks in $A^\text{ano}$ and $\text{Im}\chi_\text{loc}^\text{orb}/\pi$.

\section{Discussion and Conclusions}
\label{sec:conclusions}

Using DMFT, we have solved a model which has been previously discussed in the context of monolayer FeSe and SC induced by uniform ($\mathbf{q=0}$) nematic fluctuations. Our study provides an alternative point of view by focusing on local orbital fluctuations and their effect on superconductivity. We showed that DMFT produces a qualitatively and even quantitatively similar phase diagram to the one previously obtained by lattice QMC \cite{Dumitrescu2016}, apart from the prediction of different long-range ordered phases. In particular, DMFT predicts a relatively narrow FO phase in the strongly-correlated electron-doped regime, roughly along the line of maximum orbital fluctuations, and a CDW instability at $n=2$ and $|U| \gtrsim |U^\text{Mott}_c|\approx 4.4$. If the symmetry breaking is restricted to on-site pairing, the results are however similar, with SC most prominent on the electron doped side, near $n=3$.

For $|U| \lesssim |U^\text{Mott}_c|$ we demonstrated a clear connection between orbital fluctuations and the effective attractive interaction, which in the SC phase can be calculated from the ratio $|\mathrm{Re}\Sigma^{\mathrm{ano}}(i0^+)/\Delta|$. Both quantities peak in the same region of the phase diagram, along a line which starts near filling $n\approx 2.5$ at low $|U|$ and decreases toward $n=2$ as the interaction approaches $|U^\text{Mott}_c|$. At fixed filling, $|\mathrm{Re}\Sigma^{\mathrm{ano}}(i0^+)/\Delta|$ increases faster than $|U|$, with a correction term that scales approximately quadratically, as expected from the bubble estimate (Eq.~(\ref{Ueff})) for the effective interaction. These observations suggests a pairing induced by local orbital fluctuations, similar to the situation in repulsively interacting multiorbital systems with $J<0$, such as fulleride compounds \cite{Steiner2016,Hoshino2017}. To understand the maximum in the order parameter and $T_c$ it is however also important to consider the superfluid stiffness $D_S$, which peaks at larger dopings. Especially in the strongly correlated regime ($|U|\gtrsim |U^\text{Mott}_c|$) the order parameter reaches its largest value at $n\approx 3$ and $n\approx 1$, near the fillings corresponding to the maximum $D_S$ rather than near the peak in ${\it \Delta}\chi_\text{loc}^\text{orb}$. 

A relevant question is which parameter regime is representative of FeSe. This material is strongly correlated with repulsive Hubbard interactions within and between the $d$ orbitals. In Ref.~\onlinecite{Onari_PRL_2015} it has been argued that the coupling to local phonons can significantly screen the static interactions, leading to an overscreening of the Hund exchange, and an effective low-energy model which favors orbital fluctuations, similar to the one considered in this work. The filling per $d_{xz}$ and $d_{yz}$ orbital in monolayer FeSe is about $1.2$, according to the density functional theory plus DMFT calculation  in Ref.~\onlinecite{Moon2020}, which implies $n\approx 2.4$. A rough idea of the realistic values of $|U|$ may be obtained by comparing the computed transition temperatures $T_c$ to the experimentally established $T_c\approx 109$ K \cite{Ge_2015_NatMat_FeSe_109K}. This suggests $|U^\text{FeSe}| \lesssim 2$ ($=0.2$ eV), which places the material close to the line of maximum $\Delta \chi_\text{loc}^\text{orb}$ (black crosses in Fig~\ref{fig:fig3_colormap}(c)). 
Within the current model description, the experimentally relevant parameter regime is thus the electron-doped weak-$|U|$ region (below $U_c^\text{Mott}$), where the effective attraction is controlled by local orbital fluctuations. 

We have to note that some aspects of model (\ref{eq:model}) are debatable. The strong screening by local phonons and the resulting dominance of orbital fluctuations over spin fluctuations has been proposed in Ref.~\onlinecite{Onari_PRL_2015} in the context of the general discussion of $s_{+-}$ versus $s_{++}$ pairing in iron pnictides and the impurity effect. This work suggested an overscreening of $J$, similar to the case of A$_3$C$_{60}$, while the phonon-screened intra-orbital interaction remains positive. Model (\ref{eq:model}) also mimics a negative $J$, as mentioned in the introduction, but it also has an attractive intra-orbital interaction. This attractive $U$ may not play an essential role in the (physically relevant) weak-coupling regime, but it becomes questionable in the strong coupling regime $|U|\gtrsim |U^\text{Mott}_c|$ that was discussed in Ref.~\onlinecite{Dumitrescu2016}. 

Also, the Fermi surface structure of this model is actually for a monolayer of the bulk system \cite{Yao2009FeSemodel}, which features hole pockets at the $\Gamma$ point and electron pockets at the $X$ point in the extended Brillouin zone (1 Fe per unit cell) \cite{AnuRev2017_FeSe} for $n\sim2$.  In the \mbox{FeSe/STO} system, the hole pockets at the $\Gamma$ point sink below the Fermi energy \cite{NatCom2012_SC_FeSe}, a situation which in our model is only achieved for $n\gtrsim 3$.  However, model (\ref{eq:model}) qualitatively
captures the doping evolution of the pockets, i.e. the shrinking of the hole pockets at the $\Gamma$ point 
and the expansion of the electron pockets at the $X$ point with increasing filling.  

A recent resonant inelastic X-ray scattering study has furthermore revealed profound differences between the spin excitation spectrum of bulk and monolayer FeSe \cite{Pelliciari_2021_NC_FeSe}, which suggests a possibly important role of spin fluctuations in the pairing. Such physics is not captured by model (\ref{eq:model}).

The main purpose of the present study was to relate the concept of nematicity enhanced pairing, which has been discussed on the basis of model  (\ref{eq:model}) \cite{Yamase2013,Dumitrescu2016}, to the deeper concept of unconventional superconductivity induced by the freezing of local (spin or orbital) moments, which has emerged over the past six years \cite{Hoshino2015,Steiner2016,Werner2016,Hoshino2017,Werner2018,Yue_2021_SC}. We showed that in the realistic parameter regime, the pairing in model \eqref{eq:model} can be understood as arising from enhanced local orbital fluctuations, which grow as the system approaches an orbital-freezing regime, very similar to what occurs in A$_3$C$_{60}$ on the weak-coupling side of the $T_c$  dome. In this regime the results fit into the picture of orbital-freezing induced SC. We however also concluded that for $U>|U_c^\text{Mott}|$ and $n\approx 3$, outside the realistic regime for FeSe, the orbital-fluctuation-induced effective attraction no longer plays the dominant role in the pairing. In this regime, Ref.~\onlinecite{Dumitrescu2016} found a correlation between pairing and ${\bf q=0}$ nematic fluctuations. Our DMFT study cannot directly measure such uniform fluctuations (this would require the measurement of a vertex and a post-processing analogous to what was performed for multiorbital models in Refs.~\onlinecite{Hoshino2015,Steiner2016}). Our results however suggest that the maximum of $\Delta$ in the large-$|U|$ regime, which is dominated by the bare attraction, is primarily explained by the filling dependence of the superfluid stiffness $D_S$, which exhibits a peak near $n\approx 3$. It would be interesting to test these DMFT predictions by lattice QMC simulations.

\acknowledgements

The calculations were performed on the Beo05 cluster at the University of Fribourg, using a code based on iQIST \cite{HUANG2015140,iqist}. We acknowledge support from SNSF Grant No. 200021-196966.

\appendix

\section{Effective Interaction}
\label{App_U}
Following Ref.~\onlinecite{Hoshino2015}, we derive the effective interaction
\begin{equation}
{U}^{\text{bubble}}_{\alpha \beta}(q)=U_{\alpha \beta}-\sum_{\alpha_{1}} U_{\alpha \alpha_{1}} \chi_{\alpha_{1}}(q) {U}_{\alpha_{1} \beta}^\text{bubble}(q),
\label{eq_effective}
\end{equation}
which takes into account the effect of bubble diagrams. 
Here, $\alpha=\text{(spin,orbital)}$ is the flavor index, and $q=({\bf q},i\nu_m)$ a combined momentum and frequency index, with $\nu_m=2\pi m/\beta$ the Bosonic Matsubara frequency. 
The susceptibility $\chi_\alpha$ in the second term is defined as 
\begin{equation}
\chi_{\alpha}(q)=-\sum_{k} G_{\alpha}(k) G_{\alpha}(k+q), 
\end{equation}
with $G_{\alpha}(k)$ the single-particle Green's function for flavor $\alpha$. In the DMFT approximation, 
we only consider local vertex corrections, i.~e., $\chi_{\alpha}(q) \approx \chi_{\alpha}^{\text{loc}}(i\nu_m)$, and for local interactions
may eliminate the $q$-dependence in Eq.~\eqref{eq_effective}.
In the following, we are interested in the static limit of these local interactions, and thus use $\chi_{\alpha}^{\text{loc}}\equiv\chi_{\alpha}^{\text{loc}}(i\nu_0)=\int_0^\beta \chi_{\alpha}^{\text{loc}}(\tau)\mathrm{d}\tau$.  
In the weak-coupling limit, the above local susceptibility may be identified with either the orbital or spin susceptibility. 
As the attractive interaction $U$ increases in magnitude, the orbital susceptibility grows and the spin susceptibility is suppressed. We thus interpret 
$\chi_{\alpha}^{\text{loc}}$ as the local orbital susceptibility $\chi_\text{orb}^{\text{loc}}=\int d\tau \langle O(\tau)O(0)\rangle$ with $O=\tfrac{1}{2}(n_1-n_2)$. 
${U}^{\text{bubble}}$ may then be obtained by a matrix inversion as 
\begin{equation}
{U}^{\mathrm{bubble}}=(\mathbb{I}+U\chi^{\mathrm{orb}}_{\mathrm{loc}})^{-1}U, 
\end{equation}
where the bare interaction in matrix form (using the ordering $\left[\begin{array}{cccc}
1 & 2 & 3 & 4\\
1\uparrow & 1\downarrow & 2\uparrow & 2\downarrow
\end{array}\right]$)  reads
\begin{equation}
U=\left[\begin{array}{cccc}
0 & U & -U & -U\\
U & 0 & -U & -U\\
-U & -U & 0 & U\\
-U & -U & U & 0
\end{array}\right].
\end{equation}
The explicit calculation yields the effective static intra-orbital interaction
\begin{equation}
{U}_{1\uparrow1\downarrow}^{\mathrm{bubble}}=\frac{U}{\left[\left(1+U\chi^{\mathrm{orb}}_{\mathrm{loc}}\right)^{2}-4\left(U\chi^{\mathrm{orb}}_{\mathrm{loc}}\right)^{2}\right]}.
\end{equation}
In the weak coupling limit, we have
\begin{equation}
{U}_{1\uparrow1\downarrow}^{\mathrm{bubble}} \approx U-2U^{2}\chi^{\mathrm{orb}}_{\mathrm{loc}}+3U(U\chi^{\mathrm{orb}}_{\mathrm{loc}})^{2}.
\label{eq:Ueff_1up1dn_approx}
\end{equation}
Since $U<0$ and $\chi^{\mathrm{orb}}_{\mathrm{loc}}>0$, the bubble corrections make the intra-orbital effective interaction in Eq.~(\ref{eq:Ueff_1up1dn_approx}) more attractive. We thus expect that SC is enhanced by the local orbital fluctuations.
If we take the absolute value of the interaction and truncate Eq.~(\ref{eq:Ueff_1up1dn_approx}) at second order, we find
\begin{equation}
|{U}^\text{bubble}|=|U|+2|U|^{2}\chi^{\mathrm{orb}}_{\mathrm{loc}}.
\label{eq:Ueff_1up1dn_approx_new}
\end{equation}

\begin{widetext}

\section{Superfluid Stiffness}
\label{App_stiffness}

The stiffness, or phase rigidity, measures 
how stable the superconducting state is against phase twisting. 
In the BCS mean-field theory, $T_c$ scales with the paring gap. However, such a scaling is not valid 
in many unconventional superconductors \cite{Upper_Bounds_Tc_PRX2019}, where $T_c$ is related to the superfluid stiffness $D_S$.
Here, the superconducting order melts by fluctuations of the phase of the order parameter, rather than by the suppression of its amplitude. 
Within the framework of linear response and in the long-wave-length limit (${\bf q}\rightarrow 0$), the general formula for the stiffness \cite{Colemanbook,Simard2019} is 
\begin{equation}
D_{S}^{ab}=\frac{e^{2}}{\hbar^{2}\beta VN}\sum_{{\bf k},i\omega_{n}}\left\{ \mathrm{Tr}\left[\underline{G}({\bf k},i\omega_n)(\sigma_0 \otimes\lambda_{{\bf k}}^{b})\underline{G}({\bf k},i\omega_n)(\sigma_0\otimes\lambda_{{\bf k}}^{a})+\underline{G}({\bf k},i\omega_n)e^{i\omega_{n}0^{+}}(\sigma_3\otimes\lambda_{{\bf k}}^{ab})\right]\right\},
\label{eq:stiff}
\end{equation}
where the first and second terms of Eq.~(\ref{eq:stiff}) represent the paramagnetic and diamagnetic parts, respectively. Here 
\begin{equation}
\underline{G}({\bf k},i\omega_{n})=[i\omega_{n}\mathbb{I}_{4\times4}+\sigma_{3}\otimes\mu\mathbb{I}_{2\times2}-\sigma_{3}\otimes H_{0}({\bf k})-\Sigma^{\mathrm{Nambu}}(i\omega_{n})]^{-1}
\end{equation}
 is the interacting lattice Green's function ($4\times 4$ matrix) calculated with the local self-energy $\Sigma^{\mathrm{Nambu}}$ from DMFT. 
$\lambda_{\bf k}^a$ and $\lambda_{\bf k}^{ab}$ are the $2\times 2$ matrices
\begin{equation}
\begin{array}{c}\lambda_{{\bf k}}^{a} \equiv \partial_{{\bf k}_{a}} H_0({\bf k}),\quad \lambda_{{\bf k}}^{a b} \equiv \partial_{{\bf k}_{a}} \partial_{{\bf k}_{b}} H_0({\bf k}),\end{array}
\end{equation}
with $a,b$ an index for the Cartesian axes $x$, $y$ and $z$. 

The Kronecker product $\otimes$ in the first term of Eq.~(\ref{eq:stiff}) is 
\begin{equation}
\sigma_{0}\otimes\lambda_{{\bf k}}^{a}=\left[\begin{array}{cc}
\lambda_{{\bf k}}^{a} & 0\\
0 & \lambda_{{\bf k}}^{a}
\end{array}\right],
\end{equation}
while that in the second term corresponds to
\begin{equation}
\sigma_{3}\otimes\lambda_{{\bf k}}^{ab}=\left[\begin{array}{cc}
\lambda_{{\bf k}}^{ab} & 0\\
0 & -\lambda_{{\bf k}}^{ab}
\end{array}\right].
\end{equation}
Here, $\sigma_0=\mathbb{I}_2$ and $\sigma_3$ is the third Pauli matrix.

In the following, we list the explicit expressions for $\lambda_{\bf k}^a$ and $\lambda_{\bf k}^{ab}$:
\begin{equation}
\partial_{{\bf k}_{x}}H_{0}({\bf k})=\left(\begin{array}{cc}
4t_{3}\sin k_{x}\cos k_{y}+2t_{1}\sin k_{x} & -4t_{4}\cos k_{x}\sin k_{y}\\
-4t_{4}\cos k_{x}\sin k_{y} & 4t_{3}\sin k_{x}\cos k_{y}+2t_{2}\sin k_{x}
\end{array}\right),
\end{equation}

\begin{align}
\partial_{{\bf k}_{y}}H_{0}({\bf k}) & =\left(\begin{array}{cc}
4t_{3}\cos k_{x}\sin k_{y}+2t_{2}\sin k_{y} & -4t_{4}\sin k_{x}\cos k_{y}\\
-4t_{4}\sin k_{x}\cos k_{y} & 4t_{3}\cos k_{x}\sin k_{y}+2t_{1}\sin k_{y}
\end{array}\right),
\end{align}

\begin{align}
\partial_{{\bf k}_{x}}\partial_{{\bf k}_{x}}H_{0}({\bf k})=\left(\begin{array}{cc}
4t_{3}\cos k_{x}\cos k_{y}+2t_{1}\cos k_{x} & 4t_{4}\sin k_{x}\sin k_{y}\\
4t_{4}\sin k_{x}\sin k_{y} & 4t_{3}\cos k_{x}\cos k_{y}+2t_{2}\cos k_{x}
\end{array}\right),
\end{align}

\begin{align}
\partial_{{\bf k}_{y}}\partial_{{\bf k}_{y}}H_{0}({\bf k})=\left(\begin{array}{cc}
4t_{3}\cos k_{x}\cos k_{y}+2t_{2}\cos k_{y} & 4t_{4}\sin k_{x}\sin k_{y}\\
4t_{4}\sin k_{x}\sin k_{y} & 4t_{3}\cos k_{x}\cos k_{y}+2t_{1}\cos k_{y}
\end{array}\right),
\end{align}

\begin{align}
\partial_{{\bf k}_{x}}\partial_{{\bf k}_{y}}H_{0}({\bf k}) & =\partial_{{\bf k}_{y}}\partial_{{\bf k}_{x}}H_{0}({\bf k})=\left(\begin{array}{cc}
-4t_{3}\sin k_{x}\sin k_{y} & -4t_{4}\cos k_{x}\cos k_{y}\\
-4t_{4}\cos k_{x}\cos k_{y} & -4t_{3}\sin k_{x}\sin k_{y}
\end{array}\right).
\end{align}

\end{widetext}

\bibliography{Ref.bib}

\end{document}